%% file: wsc26paper.tex
\documentclass{wscpaperproc}
\usepackage{latexsym}
\usepackage{graphicx}
\usepackage{mathptmx}
\usepackage[T1]{fontenc}

\usepackage{amsmath}
\usepackage{amsfonts}
\usepackage{amssymb}
\usepackage{amsbsy}
\usepackage{amsthm}

\usepackage{tikz}
\usetikzlibrary{arrows.meta,patterns}
\usepackage{tabularx,booktabs}
\newcolumntype{Y}{>{\centering\arraybackslash}X}
\usepackage{diagbox}
\usepackage{multirow}
\usepackage{algorithm}
\usepackage[noend]{algpseudocode}
\usepackage[skip=0pt]{caption}
\usepackage{subcaption}
\usepackage{color,soul}
\def\R{{\mathbb{R}}}   

\def\E{{\mathbb{E}}}

\def\1{{\mathbf{1}}}

\DeclareMathOperator*{\diver}{div}

\usepackage[pdftex,colorlinks=true,urlcolor=blue,citecolor=black,anchorcolor=black,linkcolor=black]{hyperref}

\usepackage{cleveref}

\crefformat{equation}{(#2#1#3)}
\Crefname{equation}{}{}

\newtheoremstyle{wsc}
{3pt}
{3pt}
{}
{}
{\bf}
{}
{.5em}
{}

\theoremstyle{wsc}

\newtheorem{assumption}{Assumption}

\begin{document}

%
%

\pagestyle{fancyplain}

\thispagestyle{plain}
\firstPageHead{}

\chead{\fancyplain{}{\itshape Ren, Fu, and L'Ecuyer}}

\rhead{}
\cfoot{}
\renewcommand{\headrulewidth}{0pt} 

\input{wscbib.tex}           

\setlength{\baselineskip}{12.7pt}

\title{Conditional Leibniz Derivative Estimation with an Application to American Call Min-Options}


\author{\begin{center}
    Xingyu Ren\textsuperscript{1}, Michael C. Fu\textsuperscript{2}, and Pierre L'Ecuyer\textsuperscript{3}\\[11pt]
    \textsuperscript{1}Dept.~of ECE \& Inst.~for Syst.~Res., Univ.~of Maryland, College Park, MD, USA\\
    \textsuperscript{2}Robert H.~Smith Sch.~of Bus. \& Inst.~for Syst.~Res., Univ.~of Maryland, College Park, MD, USA\\
    \textsuperscript{3}DIRO, Univ.~de Montréal, Montréal, QC, Canada
\end{center}}

\maketitle

\vspace{-12pt}

\section*{ABSTRACT}
Leibniz derivative estimation is a Monte Carlo technique for estimating derivatives of a discontinuous sample performance in stochastic models with respect to parameters of interest. By combining the push-out likelihood ratio (LR) method with Leibniz integral rules, it generalizes a broad class of existing LR-based derivative estimators. However, as an LR-based method, its variance is often higher than that of perturbation analysis-based methods and may grow linearly with the dimension of the stochastic input whose distribution depends on the parameter. In this paper, we propose a recursive conditioning approach and combine it with the Leibniz derivative estimation framework. The resulting conditional Leibniz estimator does not involve LR terms and therefore is not subject to variance growth with the input dimension. It also has a simple form and is easy to implement. We apply the method to an American call min-option model, and simulation results show its effectiveness and low-variance performance.

\section{INTRODUCTION}
\label{sec:intro}
We consider the problem of estimating the derivative of the expected output performance of a stochastic system with respect to (w.r.t.) a parameter of interest. This is useful for sensitivity analysis and gradient-based optimization. The simplest method is finite-difference (FD) approximations, but it is biased and requires a bias-variance trade-off. The most fundamental unbiased direct methods are infinitesimal perturbation analysis (IPA) and the likelihood ratio (LR) method. However, IPA does not apply to discontinuous sample performance functions, and standard LR applies only to parameters of the probability distributions. A unified view of IPA and LR can be found in \citeN{l1990unified}. Various LR-based techniques have been proposed for structural parameters, including push-out LR \cite{rubinstein1992sensitivity}, support-independent unified LR-IPA (SLRIPA) \shortcite{wang2012new}, and generalized LR (GLR) \shortcite{peng2018new,peng2020generalized}. These methods rely on identifying a change of variables (push-out) that smooths the discontinuous sample performance. Such a transformation introduces the parameter into the density function, thereby causing the resulting estimator to involve LR terms. Push-out LR and SLRIPA cannot handle random inputs with parameter-dependent support. GLR can handle parameter-dependent support, but the conditions used to prove its unbiasedness are overly restrictive. Moreover, it may involve the estimation of a surface integral term that requires simulating multiple sample paths, making implementation more costly and sometimes even prohibitive. \shortciteN{ren2025stochastic} proposed a novel Leibniz framework that combines LR with the Leibniz integral rule. It includes various existing LR-based methods as special cases, allows parameter-dependent support, has shown broader applicability, and provides a single-run estimator in many cases where previous LR-based methods are either inapplicable or require multiple simulation runs. However, LR-based methods often exhibit higher variance than IPA and its variants. In particular, the variance of an LR estimator may increase linearly with the dimension of the stochastic input whose distribution depends on the parameter \cite{l1990unified}.

In this paper, we focus on the Leibniz framework of \shortciteN{ren2025stochastic} for stochastic derivative estimation of structural parameters in discontinuous sample performance functions, and combine it with a recursive conditioning approach to develop a new conditional Leibniz derivative estimator, which has the following advantages:
\begin{itemize}
    \item \textbf{Variance reduction:}  Through recursive conditioning and an alternative form of the Leibniz framework, we avoid differentiating the input density function and thereby eliminate LR terms, which in turn avoids the variance inflation often associated with LR methods for high-dimensional input.
    \item \textbf{Ease of implementation:} By eliminating LR terms, the conditional Leibniz estimator takes a simpler form than the original Leibniz derivative estimators, making it easier to implement.
\end{itemize}
Our approach is different from conditional Monte Carlo methods; see, e.g., \citeN{fu2012conditional} and \shortciteN{l2022monte}. In our approach, the primary purpose of recursive conditioning is to decompose the expectation of the sample performance. This decomposition enables recursive applications of the Leibniz framework without introducing LR terms, making the approach particularly well-suited for the sequentially structured sample performances detailed in \Cref{sec:con-leibniz,sec:simu-exmple}. The conditional Leibniz derivative estimator generalizes the method of \citeN{fu1995sensitivity}: the latter applies the univariate Leibniz integral rule and is limited to sequential decision-making models with a scalar state variable (see also \citeN{fu2026augmenting}), whereas our method applies the multivariate Leibniz integral rule and extends to models with vector-valued states. We also note that \shortciteN{peng2022variance} combine conditioning techniques with GLR methods to reduce variance. However, unlike our method, their approach does not eliminate LR terms. 

We demonstrate the advantages of the conditional Leibniz method on a two-asset American call min-option model \shortcite{detemple2003valuation}, which is a finite-horizon optimal stopping problem driven by a two-dimensional stochastic process. Specifically, we apply the conditional Leibniz method to estimate the derivative of the expected payoff w.r.t. a parameter defining the early exercise (stopping) boundary, and compare its performance with conditional GLR and FD. In this example, the resulting estimator reduces to a weighted sum of payoff functions multiplied by a conditional density, without LR terms or differentiation, which makes it particularly easy to implement.

The rest of this paper is organized as follows. In \Cref{sec:recap-leibniz}, we review Leibniz derivative estimators and explain how LR terms arise and contribute to the variance issue. In \Cref{sec:con-leibniz}, we introduce the conditional Leibniz method, including the problem formulation, assumptions, derivation, and implementation. In \Cref{sec:simu-exmple}, we apply the conditional Leibniz method to an American call min-option model and present simulation experiments. Finally, \Cref{sec:conc} concludes and discusses future research directions.

\section{A REVIEW OF LEIBNIZ DERIVATIVE ESTIMATORS}
\label{sec:recap-leibniz}
In this section, we review the Leibniz derivative estimators and explain why they can be viewed as LR-based estimators. Let $\theta\in\Theta\subset\R$ denote the parameter of interest. Consider the following sample performance:
\begin{align}\label{eq:ori-perf}
    \psi(X,\theta)=\1\{X\in h(V,\theta)\}\varphi(X,\theta),~ \theta\in\Theta,
\end{align}
under the following assumptions from \shortciteN{ren2025stochastic}.
\begin{assumption}\label{assump:leibniz}
    Let $\Theta \subset \R$ be a bounded open interval and let $V\subset \R^n$ be a Borel set. Let $X$ be an $n$-dimensional random vector supported on $\Omega \subseteq \R^n$, with a $\theta$-independent continuously differentiable density $f:\Omega\to\R$. The function $\varphi:\Omega\times\Theta\to\R$ is continuously differentiable in both arguments. The function $h:V\times\Theta\to\R^n$ is twice continuously differentiable and invertible in its first argument, and continuously differentiable in its second argument. For each $\theta$, the Jacobian of $h(\cdot,\theta)$, denoted by
    \begin{align*}
    J_h(v,\theta)=\begin{bmatrix}
        \partial_{v_1} h_1(v,\theta) &\partial_{ v_2 } h_1(v,\theta)&\cdots &\partial_{ v_n} h_1(v,\theta)\\
        \partial_{ v_1 } h_2(v,\theta)&\partial_{ v_2} h_2(v,\theta) &\cdots &\partial_{ v_n} h_2(v,\theta)\\
        \cdots &\cdots &\ddots &\cdots\\
        \partial_{ v_1} h_n(v,\theta) &\partial_{ v_2} h_n(v,\theta) &\cdots &\partial_{ v_n} h_n(v,\theta)
    \end{bmatrix},
    \end{align*}
    is $\mu$-almost everywhere (a.e.) invertible on $V$, where $\mu$ denotes the Lebesgue measure on $\R^n$.
\end{assumption}
Our goal is to estimate $\frac{d}{d\theta}\E (\psi(X,\theta))$. Under \Cref{assump:leibniz}, $\psi(X,\theta)$ is a special case of the sample performances considered in both Theorems 1 and 2 of \shortciteN{ren2025stochastic}. Therefore, both the Leibniz divergence estimator and the Leibniz integral estimator proposed there are applicable. Although this is a special case, it is still broad enough to cover a wide range of applications studied in the existing literature \shortcite{peng2018new,peng2020generalized,ren2025stochastic}, including option pricing, quantile estimation, density estimation, and control charts. Next, we present the details of both Leibniz estimators.

%

\subsection{Leibniz Integral Estimator}
Let $h^{-1}(x,\theta)$ denote the inverse of $h(v,\theta)$ w.r.t. its first argument, and define $s(x,\theta):=J_{h^{-1}}^{-1}(x,\theta)\partial_\theta h^{-1}(x,\theta)$. Under \Cref{assump:leibniz}, the change of variables $Y=h^{-1}(X,\theta)$ removes the dependence on $\theta$ from the indicator function in \Cref{eq:ori-perf}, thereby smoothing the sample performance. Applying Theorem 2 of \shortciteN{ren2025stochastic} under this change of variables yields the following expression:
\begin{align}\label{eq:ex-int-est}
    \frac{d}{d\theta}\E (\psi(X,\theta)) = \int_\Omega \psi(x,\theta)\diver(-f(x)s(x,\theta))dx
    +\int_{\partial\Omega}\psi(x,\theta)s(x,\theta)^T \vec n(x)f(x) d\sigma(x),
\end{align}
where $\diver$ is the divergence operator, defined by $\diver(\vec v)=\sum_{i=1}^n \partial_{x_i}\vec v_i$ for a vector field $\vec v:\R^n\to\R^n$ with components $\vec v_i$; $\vec n(x)$ is the outward unit normal vector on $\partial\Omega$, the boundary (surface) of $\Omega$; and $\sigma(x)$ denotes the surface measure on $\partial\Omega$, namely, the $(n-1)$-dimensional Hausdorff measure induced by the Lebesgue measure on $\R^n$.

In \Cref{eq:ex-int-est}, the first term on the right-hand side (RHS) is a volume integral over $\Omega$, whereas the second term is a surface integral over $\partial\Omega$. The surface integral may vanish under suitable conditions; for example, this occurs when $f$ vanishes on $\partial\Omega$. When it does not vanish, its estimation may require simulating multiple sample paths; see Section 5.1 in \shortciteN{ren2025stochastic} for details. Here, we focus on the volume integral term on the RHS of \Cref{eq:ex-int-est}, which can be written as
\begin{align*}
    \int_\Omega \psi(x,\theta)\diver(-f(x)s(x,\theta))dx = \E\left( \psi(X,\theta)\diver(-f(X)s(X,\theta))/f(X) \right),
\end{align*}
where the expectation $\E$ is w.r.t. the density $f$. Thus,
\begin{align}\label{eq:leibniz-int}
    -\psi(X,\theta)\diver(f(X)s(X,\theta))/f(X)
\end{align}
is an unbiased estimator. In the proof of their Theorem 2, \shortciteN{ren2025stochastic} show that, after the change of variables $Y=h^{-1}(X,\theta)$, \Cref{eq:leibniz-int} arises from differentiating the transformed density $\tilde f(y,\theta) = f(h(y,\theta)) \det|J_h(y,\theta)|$. Therefore, it is exactly an LR term.


\subsection{Leibniz Divergence Estimator}
Suppose that $h(V,\theta)\subseteq \Omega$. Applying Theorem 1 of \shortciteN{ren2025stochastic} yields the following expression:
\begin{align}\label{eq:ex-div-est}
    \frac{d}{d\theta}\E(\psi(X,\theta))
    = \int_{h(V,\theta)} \left( \partial_\theta\varphi(x,\theta) f(x)
    + \diver\left( \varphi(x,\theta) f(x)\partial_\theta h(v,\theta)\big|_{v=h^{-1}(x,\theta)} \right) \right) dx,
\end{align}
which leads to the following Leibniz divergence estimator:
\begin{align}\label{eq:leibniz-div}
   \1\{X\in h(V,\theta)\}\left(\partial_\theta\varphi(X,\theta)
    + \diver\left( \varphi(X,\theta) f(X)\partial_\theta h(v,\theta)\big|_{v=h^{-1}(X,\theta)}\right)/f(X)\right).
\end{align}
The term $\diver\left( \varphi(x,\theta) f(x) \partial_\theta h(v,\theta)|_{v=h^{-1}(x,\theta)}\right)$ has a form similar to $\diver(f(x)s(x,\theta))$ and can be interpreted similarly as an LR term.

\subsection{Elimination of LR terms}
For the Leibniz estimators in \Cref{eq:leibniz-int,eq:leibniz-div}, the LR term corresponds to the derivative of the density of $Y=h^{-1}(X,\theta)$, which has the same dimension as $X$. The LR estimator often has higher variance than IPA estimators. In particular, its variance may grow linearly with the number of stochastic inputs with parameter-dependent distributions. To reduce such variance inflation as the input dimension increases, we seek to avoid these LR terms. This can be achieved through an alternative form of the Leibniz divergence estimator. Specifically, we apply the divergence theorem to the ``LR term'' in \Cref{eq:ex-div-est}, which is exactly the divergence term as discussed earlier:
\begin{align*}
    \int_{h(V,\theta)}
    \diver\left( \varphi(x,\theta) f(x)\partial_\theta h(v,\theta)\big|_{v=h^{-1}(x,\theta)} \right)dx
    =
    \int_{\partial h(V,\theta)} \varphi(x,\theta) f(x)
    \left(\partial_\theta h(v,\theta)\big|_{v=h^{-1}(x,\theta)}\right)^{T}\vec n(x,\theta)d\sigma(x),
\end{align*}
where $\partial h(V,\theta)$ is the boundary of $h(V,\theta)$ and $\vec n(x,\theta)$ is the outward unit normal vector on $\partial h(V,\theta)$. Thus, applying the divergence theorem converts a volume integral over $h(V,\theta)$ with a pure divergence integrand into a surface integral over $\partial h(V,\theta)$ without the divergence operator. Because the divergence operator is essentially a differentiation operator, removing it also eliminates the differentiation w.r.t. the density, and hence the LR term. Consequently, \Cref{eq:ex-div-est} can be rewritten as
\begin{align}\label{eq:ex-div2-est}
    \frac{d}{d\theta}\E(\psi(X,\theta))
    = \int_{h(V,\theta)}  \partial_\theta\varphi(x,\theta) f(x) dx + \int_{\partial h(V,\theta)} 
     \varphi(x,\theta) f(x)\left(\partial_\theta h(v,\theta)\big|_{v=h^{-1}(x,\theta)}\right)^{T} \vec n(x,\theta) d\sigma(x),
\end{align}
where differentiation is taken only w.r.t. components of the sample performance, such as $\partial_\theta\varphi(x,\theta)$ and $\partial_\theta h(v,\theta)\big|_{v=h^{-1}(x,\theta)}$, rather than the density, and can therefore lead to PA-based estimators that often have lower variance. Notice that since the volume integral on the RHS of \Cref{eq:ex-int-est} does not have a pure divergence integrand, the divergence theorem cannot be applied in the same way to convert it into a surface integral and thereby eliminate the LR term.

The volume integral term on the RHS of \Cref{eq:ex-div2-est} can be written as $\int_{h(V,\theta)}  \partial_\theta\varphi(x,\theta) f(x) dx = \E (\1\{X\in h(V,\theta)\}\partial_\theta\varphi(X,\theta))$, which leads to the unbiased IPA estimator $\1\{X\in h(V,\theta)\}\partial_\theta\varphi(X,\theta)$. Estimating the surface integral in \Cref{eq:ex-div2-est} is more challenging:
\begin{itemize}
    \item Unless the surface has a simple form, such as a hyperrectangle, direct Monte Carlo sampling of an integral over $\partial h(V,\theta)$ is computationally challenging.
    \item Section 5.1 of \shortciteN{ren2025stochastic} provides a way to sample from hyperrectangular $\partial h(V,\theta)$, but the number of required sample paths grows linearly with the dimension of $X$. Each path requires sampling from the conditional density of $X$ given one of its components. Obtaining these conditional densities can be difficult unless $X$ has independent components, especially for high-dimensional or even infinite-dimensional $X$, making the estimator complex or even infeasible to implement.
\end{itemize}
To address the first issue, we note that the choice of $h$ and $V$ in the problem formulation is not unique, as long as $h(V,\theta)$ remains unchanged. When $\partial h(V,\theta)$ is not hyperrectangular, one may seek an alternative formulation of $h$ and $V$ such that $V$ is a hyperrectangle, and then apply a change of variables (aka pullback) to transform the surface integral over $\partial h(V,\theta)$ into one over $\partial V$. Such a formulation with hyperrectangular $V$ is achievable in many sequential decision models; see, e.g., Example 4 of \shortciteN{ren2025stochastic}. To address the second issue, we can apply conditioning techniques. By recursively conditioning on suitable components of the vector $X$, we can not only reduce variance directly, but also reduce the problem dimension and thereby make the estimator easier to implement.

In the rest of this paper, we focus on the case where $\partial h(V,\theta)$ is hyperrectangular and apply the approach in \shortciteN{ren2025stochastic} to estimate the surface integral in \Cref{eq:ex-div2-est}, with the goal of developing a derivative estimator that is simple to implement and has low variance. The case in which $\partial h(V,\theta)$ is not hyperrectangular and requires an additional change of variables is left for future research.

\section{CONDITIONAL LEIBNIZ DERIVATIVE ESTIMATOR}
\label{sec:con-leibniz}
In this section, we consider the input $X$ as a finite-horizon multidimensional stochastic process, with $h(V,\theta)$ given by a product of subsets corresponding to the process over time. By sequentially conditioning on the state of the process at each time period, we decompose the original derivative estimation problem into a collection of lower-dimensional subproblems and iteratively apply \Cref{eq:ex-div2-est} to construct the conditional Leibniz estimator.

Specifically, let $X:=\{X_t\}_{t=1}^N$ be a stochastic process, where $X_t=(X_{t,1},\cdots,X_{t,n})$ is an $n$-dimensional random vector for each $t$. We consider the same sample performance $\psi(X,\theta)=\1\{X\in h(V,\theta)\}\varphi(X)$ as in \Cref{eq:ori-perf} under \Cref{assump:leibniz}. We assume that $\varphi(X)$ does not depend on $\theta$. If instead $\varphi(X,\theta)$ depends on $\theta$ and is differentiable in $\theta$ as in \Cref{assump:leibniz}, then we only need to augment the conditional Leibniz estimator developed in this section with the additional IPA term $\partial_\theta \varphi(X,\theta)$ to obtain an unbiased estimator. We impose the following additional assumption.
\begin{assumption}\label{assump:con-leibniz-1}
    There exist functions $\{h_t\}_{t=1}^N$ and sets $\{V_t\}_{t=1}^N$ such that $h(V,\theta)=h_1(V_1,\theta)\times \cdots \times h_N(V_N,\theta)$, where, for each $t=1,\cdots,N$, $h_t:V_t\times\Theta\to\R^n$ is twice continuously differentiable and invertible in its first argument, and continuously differentiable in its second argument, with $V_t\subseteq\R^n$.
\end{assumption}
Under \Cref{assump:con-leibniz-1}, we can write:
\begin{align*}
    \psi(X,\theta)=\varphi (X)\prod_{i=1}^{N}\1\{X_i \in h_i(V_i,\theta)\} .
\end{align*}
\Cref{assump:con-leibniz-1} is naturally satisfied in sequential decision models whose decision rules can be characterized by a collection of decision regions $\{h_t(V_t,\theta)\}_{t=1}^N$ over time, such as American option pricing and control charts. In the option pricing setting, $X$ may represent the price process, $\varphi(X)$ may represent the payoff function, and $h_t(V_t,\theta)$ represents the early exercise region at time $t$. In the control chart setting, $X$ may represent the manufacturing process being monitored, while $h_t(V_t,\theta)$ represents the desired region within which the process should remain. In both examples, $\theta$ can be a decision parameter that controls the shape of the regions $h_t(V_t,\theta)$.

Viewing $X$ as a stochastic process and $\{h_t(V_t,\theta)\}_{t=1}^N$ as the decision regions of the corresponding sequential decision problem, we can outline a sequential conditioning Leibniz method based on \Cref{eq:ex-div2-est} as follows. We begin at $t=1$ and condition on $X_1$, which allows us to write $\E(\psi(X,\theta))$ as an integral over $h_1(V_1,\theta)$ w.r.t. $x_1$, where the integrand is the conditional expectation of the sample performance given $X_1=x_1$. Applying \Cref{eq:ex-div2-est} yields two terms: a surface integral over $\partial h_1(V_1,\theta)$ and a volume integral whose integrand is the derivative of the conditional sample performance given $X_1=x_1$. We then apply the same procedure recursively to the derivative appearing in the volume integral: at time $t=2$, we condition on $X_2$, rewrite the corresponding conditional sample performance as an integral over $h_2(V_2,\theta)$ w.r.t. $x_2$, and apply \Cref{eq:ex-div2-est} again. Continuing in this way for $t=1,\dots,N$, we obtain a representation of $\frac{d}{d\theta}\E(\psi(X,\theta))$ as a sum of surface integrals over $\{\partial h_t(V_t,\theta)\}_{t=1}^N$. If each $h_t(V_t,\theta)$ is rectangular, a condition that will be imposed explicitly later, then these surface integrals can be estimated using the method in Section 5.1 of \shortciteN{ren2025stochastic}. Since this procedure only involves repeated applications of \Cref{eq:ex-div2-est}, it does not generate any LR term.

We now formally derive the conditional Leibniz estimator. We begin by conditioning on $X_1$:
\begin{align*}
    \E (\psi(X,\theta)) &= \E\left( \1\{X_1\in h_1(V_1,\theta)\} \E\left( \prod_{i=2}^{N}\1\{X_i \in h_i(V_i,\theta)\} \varphi (X) \bigg| X_1 \right)  \right)\\
    &=\int_{h_1(V_1,\theta)} \E\left( \prod_{i=2}^{N}\1\{X_i \in h_i(V_i,\theta)\} \varphi (X) \bigg| X_1=x_1 \right) f_{X_1}(x_1) dx_1,
\end{align*}
where $f_{X_1}$ is the marginal density of $X_1$. Applying \Cref{eq:ex-div2-est} to this integral gives
\begin{align}\label{eq:con-leib-step1}
    \begin{split}
        \frac{d}{d\theta}\E (\psi(X,\theta)) &= \E\left( \1\{X_1\in h_1(V_1,\theta)\} \frac{d}{d\theta}\E\left( \prod_{i=2}^{N}\1\{X_i \in h_i(V_i,\theta)\} \varphi (X) \bigg| X_1 \right)  \right)\\
        &+\int_{\partial h_1(V_1,\theta)} \E\left( \prod_{i=2}^{N}\1\{X_i \in h_i(V_i,\theta)\} \varphi (X) \bigg| X_1=x_1 \right) f_{X_1}(x_1) \vec v_1(x_1,\theta)^T \vec n_1(x_1,\theta)  d\sigma(x_1),
    \end{split}
\end{align}
where $\vec v_t(x_t,\theta):=\partial_\theta h_t(v_t,\theta)\big|_{v_t=h_t^{-1}(x_t,\theta)}$, $\vec n_t(x_t,\theta)$ is the outward unit normal vector on $\partial h_t(V_t,\theta)$, and $\sigma(x_t)$ is the surface measure on $\partial h_t(V_t,\theta)$.

We defer estimation of the surface integral over $\partial h_1(V_1,\theta)$ in \Cref{eq:con-leib-step1} and first focus on the expectation term on the RHS. Define $X_{t_1:t_2}:=(X_{t_1},\cdots,X_{t_2})$ for $t_1\leq t_2$. For any random vectors $Y$ and $Z$, let $f_{Y|Z}(y|z)$ denote the conditional density of $Y=y$ given $Z=z$. To evaluate the derivative of the conditional expectation given $X_1$, we condition on $X_2$, rewrite the conditional expectation as an integral w.r.t. $x_2$, and apply \Cref{eq:ex-div2-est}:
\begin{align*}
    &\frac{d}{d\theta}\E\left(  \prod_{i=2}^{N}\1\{X_i \in h_i(V_i,\theta)\} \varphi (X) \bigg| X_1 \right)\\ 
    &= \E\left( \1\{X_2\in h_2(V_2,\theta)\} \frac{d}{d\theta}\E\left( \prod_{i=3}^{N}\1\{X_i \in h_i(V_i,\theta)\} \varphi (X) \bigg| X_2,X_1 \right) \bigg| X_1 \right)\\
    &+\int_{\partial h_2(V_2,\theta)} \E\left( \prod_{i=3}^{N}\1\{X_i \in h_i(V_i,\theta)\} \varphi (X) \bigg| X_2=x_2, X_1 \right) f_{X_2|X_1}(x_2|X_1) \vec v_2(x_2,\theta)^T \vec n_2(x_2,\theta)  d\sigma(x_2).
\end{align*}
Repeating this argument to evaluate the derivatives of the conditional expectations sequentially over all time steps yields the following expression, involving only surface integrals over ${\partial h_t(V_t,\theta)}_{t=1}^N$:
\begin{align}\label{eq:sur-int}
    \begin{split}
        &\frac{d}{d\theta}\E (\psi(X,\theta))=\sum_{t=1}^N\E\Bigg(\prod_{k=1}^{t-1}\1\{X_k\in h_k(V_k,\theta)\}\\ 
        &\int_{\partial h_t(V_t,\theta)} \E\left( \prod_{j=t+1}^{N}\1\{X_j \in h_j(V_j,\theta)\} \varphi (X) \bigg| X_t=x_t, X_{1:t-1}\right)
        f_{X_t | X_{1:t-1}}(x_t | X_{1:t-1})
    \vec v_t(x_t,\theta)^T\vec n_t(x_t,\theta) d\sigma(x_t)  \Bigg).
    \end{split}
\end{align}

We now apply the method in Section 5.1 of \shortciteN{ren2025stochastic} to estimate these surface integrals in the case where each $\partial h_t(V_t,\theta)$ is hyperrectangular, as assumed below.
\begin{assumption}\label{assump:con-leibniz-2}
    For each $t$, the set $h_t(V_t,\theta)$ is a hyperrectangle in $\R^n$ of the form $h_t(V_t,\theta)=[a_{t,1},b_{t,1}]\times\cdots\times[a_{t,n},b_{t,n}]$. Its boundary consists of its $2n$ faces and can be written as $\partial h_t (V_t,\theta) = \cup_{i=1}^{n}\left(\partial h_{a_{t,i}}\cup\partial h_{b_{t,i}}\right)$, where $\partial h_{x_{t,i}}: = [a_{t,1},b_{t,1}]\times\cdots\times\{x_{t,i}\}\times\cdots\times[a_{t,n},b_{t,n}],~x_{t,i}=a_{t,i} \text{ or }b_{t,i}$.
\end{assumption} 
For each $i=1,\cdots,n$, the outward unit normal vectors on the faces $\partial h_{a_{t,i}}$ and $\partial h_{b_{t,i}}$ are $-e_i$ and $e_i$, respectively, where $e_i$ is the $i^{\text{th}}$ standard basis vector in $\R^n$. Let $x_{t,-i}$ denote the vector obtained by removing the $i^{\text{th}}$ coordinate from $x_t$. Then the surface integral over $\partial h_t(V_t,\theta)$ can be expressed as the sum of $(n-1)$-dimensional volume integrals over the $2n$ faces as follows:
\begin{align}\label{eq:sur-int}
    \begin{split}
        &\int_{\partial h_t(V_t,\theta)}
    \E\left(
        \prod_{j=t+1}^{N}\1\{X_j\in h_j(V_j,\theta)\}\varphi(X)
        \bigg| X_t=x_t, X_{1:t-1}
    \right)
    f_{X_t| X_{1:t-1}}(x_t| X_{1:t-1}) 
    \vec v_t(x_t,\theta)^T\vec n_t(x_t,\theta) d\sigma(x_t) \\
    &=
    \sum_{i=1}^{n}
    \int_{\partial h_{x_{t,i}}}
    \E\left(
        \prod_{j=t+1}^{N}\1\{X_j\in h_j(V_j,\theta)\}\varphi(X)
        \bigg| X_t=x_t, X_{1:t-1}
    \right)
    f_{X_t| X_{1:t-1}}(x_t| X_{1:t-1}) 
    \vec v_t(x_t,\theta)^Te_i d\sigma(x_t)
    \bigg|_{x_{t,i}=a_{t,i}}^{x_{t,i}=b_{t,i}}.
    \end{split}
\end{align}
Let $\prod_{j \ne i} [a_{t,j}, b_{t,j}]$ denote the projection of $\partial h_{x_{t,i}}$ onto the coordinates other than the $i^{\text{th}}$ one. Then each volume integral over $\partial h_{x_{t,i}}$ can be further expressed as a conditional expectation, as follows (taking $x_{t,i}=a_{t,i}$ as an example):
\begin{align}
    &\quad \int_{\prod_{j \ne i} [a_{t,j}, b_{t,j}]}
    \E\left(
        \prod_{j=t+1}^{N}\1\{X_j\in h_j(V_j,\theta)\}\varphi(X)
        \bigg| X_t=x_t, X_{1:t-1}
    \right)
    f_{X_t| X_{1:t-1}}(x_t| X_{1:t-1}) 
    \vec v_t(x_t,\theta)^Te_i dx_{t,-i} \bigg|_{x_{t,i}=a_{t,i}} \nonumber
    \\
    &=
    \int_{\prod_{j \ne i} [a_{t,j}, b_{t,j}]}
    \E\left(
        \prod_{j=t+1}^{N}\1\{X_j\in h_j(V_j,\theta)\}\varphi(X) \vec v_t(x_t,\theta)^Te_i
        \bigg| X_{t,i}=a_{t,i}, X_{t,-i}=x_{t,-i}, X_{1:t-1}
    \right) \nonumber\\
    &\quad
    \times f_{X_{t,-i}|  X_{t,i}, X_{1:t-1}}(x_{t,-i}| X_{t,i}=a_{t,i}, X_{1:t-1}) 
     dx_{t,-i} \times f_{X_{t,i}|   X_{1:t-1}}(a_{t,i}| X_{1:t-1}) \nonumber
    \\
    &=
    \E\left(
        \prod_{j=t+1}^{N}\1\{X_j\in h_j(V_j,\theta)\}\varphi(X) v_t(X_t,\theta)^Te_i
        \bigg| X_{t,i}=a_{t,i}, X_{t-1},\cdots,X_1
    \right)
    f_{X_{t,i}|   X_{1:t-1}}(a_{t,i}| X_{1:t-1}). \label{eq:ind-sur-int}
\end{align}
To estimate the quantity in the last line of \Cref{eq:ind-sur-int}, we can perform Monte Carlo sampling as follows: first generate $X_{1:t-1}$; then, conditional on $X_{t,i}=a_{t,i}$, generate $X_{t,-i}$ and then the remaining states $X_{t+1:N}$. This yields a complete sample path $X_{1:N}$, from which we can evaluate both the conditional expectation term and the conditional density term. The same applies to estimate the integral over the face $\partial h_{b_{t,i}}$. 

Let $N_0=\min\{t\leq N | X_t \notin h_t(V_t,\theta) \}$, with the convention that $N_0=\infty$ if $X_t\in h_t(V_t,\theta)$ for all $t=1,\cdots,N$. To construct an estimator based on \Cref{eq:sur-int}, we need to estimate the unconditional expectations on the RHS and sum them. For each expectation, suppose we use Monte Carlo sampling to generate a path $X_{1:N}$. If $t>N_0$, then $\prod_{k=1}^{t-1}\1\{X_k\in h_k(V_k,\theta)\}=0$, so this sample makes no contribution to the estimation of that expectation. Therefore, we only need to consider $t=1,\cdots,\min(N_0,N)$. For such $t$, we need to estimate the surface integral over $\partial h_t(V_t,\theta)$, for which the sampling scheme described for \Cref{eq:sur-int,eq:ind-sur-int} applies. \Cref{alg:sur-int} gives the sampling procedure for a single Monte Carlo replication used to estimate $\frac{d}{d\theta}\E (\psi(X,\theta))$, and the final estimate is obtained by averaging independent replications.
\begin{algorithm}
\caption{A single replication of the conditional Leibniz estimator for $\frac{d}{d\theta}\E (\psi(X,\theta))$}
\label{alg:sur-int}
\begin{algorithmic}[1]
\State Generate a sample path $\{\tilde X_t\}$ sequentially until time $\min(N_0,N)$. 
\For{$t=1,\cdots,\min(N_0,N)$}
    \For{$i=1,\cdots,n$}
        \For{$x_{t,i}\in\{a_{t,i},\,b_{t,i}\}$}
            \State Sample $X_{t,-i}$ from its conditional distribution given $(\tilde X_{1:t-1},X_{t,i}=x_{t,i})$.
            \State Sample $X_{t+1:N}$ sequentially.
            \If{$X_j\in h_j(V_j,\theta)$ for all $j=t+1,\cdots,N$}
                \State Set $\mathcal S (x_{t,i})
                =
                \varphi(\tilde X_{1:t-1},X_{t:N})\vec v_t(X_t,\theta)^Te_i
                f_{X_{t,i}|X_{1:t-1}}(x_{t,i}|\tilde X_{1:t-1})$.
            \Else
                \State Set $\mathcal S (x_{t,i})=0$.
            \EndIf
        \EndFor
    \EndFor
\EndFor
\State \Return $\hat D = \sum_{t=1}^{\min(N_0,N)}\sum_{i=1}^{n}\big(\mathcal S(b_{t,i})-\mathcal S(a_{t,i})\big)$.
\end{algorithmic}
\end{algorithm}

\section{SIMULATION EXAMPLE}
\label{sec:simu-exmple}
In this section, we consider an American call min-option model driven by a two-dimensional price process \shortcite{detemple2003valuation}, where the payoff upon exercise is based on the minimum of the two components of the price vector. \shortciteN{detemple2003valuation} characterized the optimal decision rule by a time-dependent early exercise (optimal stopping) region: once the price vector enters this region, the option is exercised. They also showed that this region can be reasonably approximated by a rectangle, which is easy to parametrize. We apply the conditional Leibniz estimator to estimate the derivative of the expected payoff w.r.t. the parameter defining the early exercise region.

\begin{figure}[bt]
\caption{Illustration of the optimal early exercise region in a two-dimensional American call min-option model. The optimal early exercise region consists of the gray area together with the larger red shaded rectangular area, so the red rectangle is taken as an approximation of the optimal early exercise region.}\label{fig:early-region}
\centering
\begin{tikzpicture}[scale=0.6,>=Stealth,line cap=round,line join=round]

\def\xmax{8.2}
\def\ymax{8.2}
\def\L{4.15}

\coordinate (C) at (2.5,2.5);              
\coordinate (U) at (\L,5.8);           
\coordinate (R) at (5.8,\L);           

\draw[->, very thick] (0,0) -- (\xmax,0) node[right] {$S_{\tau,1}$};
\draw[->, very thick] (0,0) -- (0,\ymax) node[above] {$S_{\tau,2}$};
\node[below left] at (0,0) {$O$};
\draw[dashed, thick] (0,0) -- (6.7,6.7);


\fill[gray!22] (\L,\L) rectangle (\xmax,\ymax);

\fill[gray!22]
(C)
.. controls (3.10,3.55) and (\L,5.10) .. (U)
-- (\L,\L)
-- (R)
.. controls (5.10,\L) and (3.55,3.10) .. (C)
-- cycle;

\draw[black, thick]
(C) .. controls (3.30,3.75) and (\L,5.10) .. (U);
\draw[black, thick] 
(U) -- (\L,7.8);

\draw[black, thick]
(C) .. controls (3.75,3.30) and (5.10,\L) .. (R);
\draw[black, thick] 
(R) -- (7.8,\L);

\fill[
    pattern=north east lines,
    pattern color=red!75!black
] (\L,\L) rectangle (\xmax,\ymax);

\draw[red!75!black, dashed, thick] (\L,0) -- (\L,8.2);
\draw[red!75!black, dashed, thick] (0,\L) -- (8.2,\L);
\node[below, red!75!black] at (\L,0) {$L_\tau$};
\node[left,  red!75!black] at (0,\L) {$L_\tau$};

\node[fill=white, inner sep=1.5pt] at (5.85,6.15) {$h_\tau(V_\tau,L_\tau)$};

\end{tikzpicture}
\end{figure}
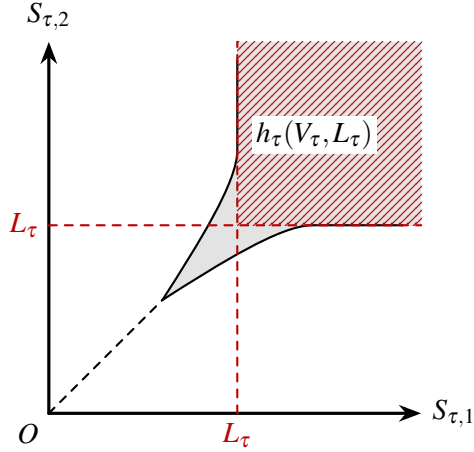

Suppose the price process, denoted by $S_t=(S_{t,1},S_{t,2})$, follows a two-dimensional geometric Brownian motion:
\begin{align*}
    d S_{t,i} = S_{t,i} \left( r_i dt + \sigma_i d W_{t,i} \right),~i=1,2.
\end{align*}
where $r_i\in\R$ is the drift rate of the $i$th asset, $\sigma_i\in\R$ is its volatility parameter, and $W_{t,i}$ is a standard Brownian motion for each $i$. We discretize the process over a time horizon $T$ divided into $N$ steps, with step size $\Delta t = T/N$. The discrete-time dynamics are given by
\begin{align}\label{eq:dynamics}
     S_{t+1,i}=S_{t,i} \exp\{ (r_i-\sigma_i^2/2)\Delta t + \sigma_i \sqrt{\Delta t} Z_{t,i} \},~i=1,2,~t=0,\dots,N-1,
\end{align}
where $\{Z_{t,i}\}$ are standard normal random variables. The two processes $\{Z_{t,1}\}$ and $\{Z_{t,2}\}$ are each independent and identically distributed but can be correlated with each other. Let $K>0$ be the strike price. If the option is exercised at step $t$, the corresponding discounted payoff is
\begin{align*}
    J_t(S_t)=  \exp\{-r_0 t \Delta t\} (S_{t,1}\wedge S_{t,2} -K)^+.
\end{align*}
where $r_0$ is the risk-free rate. The sequential decision problem is as follows: at each step $t$, if we exercise the option, we receive the payoff $J_t(S_t)$ and the decision process terminates; otherwise, we continue, with exercise being mandatory at step $N$. \shortciteN{detemple2003valuation} showed that for each step $t$, the optimal early exercise region consists of a rectangle along the diagonal together with a spike along the diagonal pointing toward the origin, as illustrated by the gray shaded area in \Cref{fig:early-region}. They also showed that the rectangular part provides a reasonable approximation; see the red shaded region in \Cref{fig:early-region}. Accordingly, in this section, we consider rectangular early exercise regions.

For simplicity, we assume symmetric parameters $r_0=r_1=r_2=:r$, $\sigma_1=\sigma_2=:\sigma$, and identical initial prices for the two assets, $S_{0,1}=S_{0,2}$. Moreover, we assume that the two components of the price process, $\{S_{t,1}\}$ and $\{S_{t,2}\}$, are independent. Under these conditions, the rectangular part of the optimal early exercise region is a square. Therefore, at each step $t$, we consider an early exercise region of the form $\{s_t\in\R_+^2 \mid s_{t,1}>L_t,\ s_{t,2}>L_t\}$, where $L_t\in\R_+$ is the only decision parameter characterizing the early exercise region. To apply \Cref{alg:sur-int}, we need to express this region in the form $h_t(V_t,L_t)$ for some set $V_t\subseteq\R^2$ and some function $h_t:\R^2\times\R_+\to\R^2$ satisfying \Cref{assump:con-leibniz-2}. Using the parametrization method in Example 4 of \shortciteN{ren2025stochastic}, we write $\{s_t\in\R_+^2 \ | \ s_{t,1}>L_t,~ s_{t,2}>L_t\}:=h_t(V_t,L_t)$, with $V_t=(0,\infty)^2$, $h_t=(h_{t,1},h_{t,2})$, and $h_{t,i}(v,L_t)=v_i+L_t$ for $i=1,2$.

Given the early exercise regions $\{h_t(V_t,L_t)\}_{t=1}^N$, the final payoff of the American call min-option, i.e., the sample performance, can be written as
\begin{align*}
    \psi(S_{1:N}) = \sum_{t=1}^N \psi_t (S_{1:t}),
\end{align*}
where
\begin{align*}
    &\psi_t (S_{1:t}) = \left( \prod_{k=1}^{t-1} \1\{S_k\notin h_k(V_k,L_k)\} \right) \1\{S_t\in h_t(V_t,L_t)\} J_t(S_t),~t<N, \\
    &\psi_N (S_{1:N}) = \prod_{k=1}^{N-1} \1\{S_k\notin h_k(V_k,L_k)\} J_N(S_N).
\end{align*}

We are interested in estimating the derivative of $\E(\psi(S_{1:N}))$ w.r.t. the boundary parameter $L_\tau$ for some fixed step $\tau\leq N-1$. To do so, we apply \Cref{alg:sur-int} to each $\psi_t(S_{1:t})$, $t=1,\cdots,N$, and then aggregate the resulting estimators to estimate the derivative of $\E(\psi(S_{1:N}))$. Specifically, for $t<\tau$, $\psi_t(S_{1:t})$ does not depend on $L_\tau$, so $\frac{d}{dL_\tau}\E(\psi_t(S_{1:t}))=0$. For each term $\psi_t(S_{1:t})$ with $t> \tau$, we rewrite it in a form consistent with \Cref{eq:ori-perf} and \Cref{assump:con-leibniz-1} in order to apply \Cref{alg:sur-int}, as follows:
\begin{align}
        \psi_t (S_{1:t}) &= \left( \prod_{k=1,k\neq \tau}^{t-1} \1\{S_k\notin h_k(V_k,L_k)\} \right) \left(1-\1\{S_\tau\in h_\tau(V_\tau,L_\tau)\}\right) \1\{S_t\in h_t(V_t,L_t)\}J_t(S_t)\nonumber\\
        &= \left( \prod_{k=1,k\neq \tau}^{t-1} \1\{S_k\notin h_k(V_k,L_k)\} \right)J_t(S_t)\1\{S_t\in h_t(V_t,L_t)\} \\
        &- \left( \prod_{k=1,k\neq \tau}^{t-1} \1\{S_k\notin h_k(V_k,L_k)\} \right) \1\{S_\tau\in h_\tau(V_\tau,L_\tau)\} \1\{S_t\in h_t(V_t,L_t)\}J_t(S_t). \label{eq:leib-ind-ame}
\end{align}
Since only the term in \Cref{eq:leib-ind-ame} contains $L_\tau$, it is the only term that contributes to the derivative w.r.t. $L_\tau$. Therefore, for $t>\tau$, we may equivalently redefine $\psi_t(S_{1:t})$ using only that term and ignore the term on the second line. A few additional technical details should be noted in the implementation of \Cref{alg:sur-int}:
\begin{itemize}
    \item  For each $\psi_t(S_{1:t})$, $t\ge \tau$, the parameter $L_\tau$ appears only through the single indicator $\1\{S_\tau\notin h_\tau(V_\tau,L_\tau)\}$. Therefore, when applying \Cref{alg:sur-int}, only the surface integral over $\partial h_\tau(V_\tau,L_\tau)$ needs to be evaluated. That is, in line 2 of \Cref{alg:sur-int}, we only need to consider the case $t=\tau$.
    
    Note that if all the boundary parameters $L_t$ are parameterized by a common parameter, say $\theta\in\R$, then the derivative w.r.t. $\theta$ can be estimated by aggregating the individual estimators for each $L_t$ and multiplying them by $\partial L_t/\partial\theta$; see a similar idea in \citeN{heidergott2001option}. In that case, the surface integral at each step $t$ contributes to the derivative. We leave this as a direction for future research.
    \item Notice that two of the four edges of $h_\tau(V_\tau,L_\tau)=(L_\tau,\infty)^2$ lie at $s_{\tau,1}=\infty$ and $s_{\tau,2}=\infty$. Since the density of $S_\tau$ vanishes at infinity, only the other two edges, namely those on $s_{\tau,1}=L_\tau$ and $s_{\tau,2}=L_\tau$, contribute to the derivative estimate, i.e., in line 4 of \Cref{alg:sur-int}, we only need to consider the case $x_{t,i}=a_{t,i}=L_\tau$.
    \item Since the price process $\{S_t\}$ is Markovian, in implementing line 8 of \Cref{alg:sur-int}, the conditional density given the entire past history reduces to the corresponding Markov transition density for \Cref{eq:dynamics}. Specifically, we only need to consider $f_{S_{\tau,i}| S_{\tau-1}}(\cdot| S_{\tau-1})$, $i=1,2$, the conditional density of $S_{\tau,i}$ given $S_{\tau-1}$, which is given by
    \begin{align*}
        f_{S_{\tau,i}|S_{\tau-1}}(s|S_{\tau-1})=\phi \left(\left(\log \frac{s}{S_{\tau-1,i}}-\left(r-\frac{\sigma^2}{2}\right)\Delta t\right)\frac{1}{\sigma \sqrt{\Delta t}}\right)\frac{1}{\sigma s \sqrt{\Delta t}},
    \end{align*}
    where $\phi$ denotes the standard normal density.
\end{itemize}
Let $\tau_0:=\min\{t\leq N-1 | S_t\in h_t(V_t,L_t)\}$, with the convention that $\tau_0=N$ if the set is empty. \Cref{alg:example-leibniz} describes the sampling procedure for one Monte Carlo replication for estimating $\frac{d}{d L_\tau}\E (\psi(S_{1:N}))$.

\begin{algorithm}
\caption{A single replication of the conditional Leibniz estimator for $\frac{d}{d L_\tau}\E (\psi(S_{1:N}))$}\label{alg:example-leibniz}
\begin{algorithmic}[1]
\State \textbf{input} fixed $\tau\in\{1,\cdots,N-1\}$.
\For{$t=1,\cdots,\tau-1$}
\If{$S_t\in h_t(V_t,L_t)$}
    \State \textbf{return} $0$ and \textbf{stop}.
\EndIf
\EndFor

\State Sample the price process at $\tau$: $S_\tau=(s_{\tau,1},s_{\tau,2})$.

\If{$s_{\tau,2}<L_\tau$}
    \State $\mathcal S_1 = 0$.
\Else
    \State Set $S_\tau=(L_\tau,s_{\tau,2})$, and then sample $\tau_0$ and the corresponding $S_{\tau_0}$.
    \State $\mathcal S_1 = \big( J_{\tau_0}(S_{\tau_0}) - J_\tau(L_\tau,s_{\tau,2}) \big) f_{S_{\tau,1}|S_{\tau-1}}(L_\tau|S_{\tau-1})$.
\EndIf

\If{$s_{\tau,1}<L_\tau$}
    \State $\mathcal S_2 = 0$.
\Else
    \State Set $S_\tau=(s_{\tau,1},L_\tau)$, and then sample $\tau_0$ and the corresponding $S_{\tau_0}$.
    \State $\mathcal S_2 = \big( J_{\tau_0}(S_{\tau_0}) - J_\tau(s_{\tau,1},L_\tau) \big) f_{S_{\tau,2}|S_{\tau-1}}(L_\tau|S_{\tau-1})$.
\EndIf

\State \Return $\hat D = \mathcal S_1 + \mathcal S_2$.
\end{algorithmic}
\end{algorithm}

We compare the conditional Leibniz method with the conditional GLR method of \shortciteN{peng2022variance}, outlined as follows. To implement conditional GLR, we condition on all input variables other than $S_\tau$, namely $\{S_t\}_{t=1}^N \setminus \{S\tau\}$, and set $g(S_\tau,L_\tau)=(S_{\tau,1}/L_\tau,S_{\tau,2}/L_\tau)$ for the change of variables in Section 2 of \shortciteN{peng2022variance}. Let $\mathcal{T}:=\min\{t< N | S_t\in h_t(V_t,L_t)\}\wedge N$ denote the early exercise step. The resulting conditional GLR estimator is then given by
\begin{align*}
    &\quad\1\{\mathcal T=\tau\}\left(J_\tau(S_\tau)d_0(S_{\tau-1:\tau},L_\tau)+\exp(-r\tau\Delta t)\1\{S_{\tau,1}\wedge S_{\tau,2} -K\geq 0\}\left(S_{\tau,1}\wedge S_{\tau,2}\right)\right)/L_\tau\\
    &+\1\{\mathcal T>\tau\}J_\tau(S_\tau)d_1(S_{\tau-1:\tau+1},L_\tau),
\end{align*}
where
\vspace*{-1em}
\begin{align*}
    &d_0(s_{\tau-1:\tau},L_\tau)=\sum_{i=1}^{2}\left(\partial_{s_{\tau,i}}\log f_{S_{\tau,i}|S_{\tau-1}}(s_{\tau,i}|s_{\tau-1})
    +1/s_{\tau,i}
    \right)s_{\tau,i}/L_\tau,\\
    &d_1(s_{\tau-1:\tau+1},L_\tau)=\sum_{i=1}^{2}\left(\partial_{s_{\tau,i}}\log f_{S_{\tau,i}|S_{\tau-1}}(s_{\tau,i}|s_{\tau-1})
    +\partial_{s_{\tau,i}}\log f_{S_{\tau+1,i}|S_{\tau}}(s_{\tau+1,i}|s_{\tau})
    +1/s_{\tau,i}
    \right)s_{\tau,i}/L_\tau.
\end{align*}

In the simulation experiment, we set $S_{0,1}=S_{0,2}=5$, $T=2$, $\tau=2$, $r=0.1$, $\sigma=0.3$, and $K=3$. The boundary parameters $\{L_t\}$ are specified by the linear function $L_t=a+b(t-1)$ with $a=5.5$. We consider four combinations of $(N,b)\in\{(5,0.5),(10,0.5),(10,0.3),(15,0.1)\}$ and compare the following estimators: the conditional Leibniz estimator, the conditional GLR estimator, and FD estimators with common random numbers using perturbation sizes $0.1$, $0.01$, and $0.001$. Each estimator is implemented using $10^6$ independent replications. The results are summarized in Table \ref{tab:sim_results}, where each point estimate is reported together with its standard error in parentheses. We observe that the conditional Leibniz estimator has substantially lower variance than the other estimators. Moreover, its performance is not sensitive to the dimension of the stochastic input, which is $2N$, whereas the variance of the conditional GLR estimator shows an obvious increasing trend as $N$ increases. For the FD estimators, none of the perturbation sizes yields a bias--variance trade-off that makes their performance comparable to that of the conditional Leibniz method.
\begin{table}[h]
\centering
\caption{Simulation results based on $10^6$ independent replications, with $S_{0,1}=S_{0,2}=5$, $T=2$, $\tau=2$, $r=0.1$, $\sigma=0.3$, and $K=3$. Values are point estimates of $\frac{d}{dL_\tau}\E(\psi(S_{1:N}))$, with standard errors in parentheses.}
\label{tab:sim_results}
\vspace*{0.1em}
\begin{tabular}{l c c c c}
\toprule \multirow{2}{*}{\textbf{Estimator}} & \multicolumn{4}{c}{\textbf{$(N,b)$}} \\ \cmidrule(lr){2-5} & \textbf{$(5,0.5)$} & \textbf{$(10,0.5)$} & \textbf{$(10,0.3)$} & \textbf{$(15,0.1)$} \\ \midrule
Conditional Leibniz         & -0.0245 (0.0002) & -0.0266 (0.0002) & -0.0329 (0.0003) & -0.0352 (0.0003) \\
Conditional GLR             & -0.0238 (0.0029) & -0.0225 (0.0045) & -0.0311 (0.0046) & -0.0347 (0.0060) \\
FD ($0.1$)                  & -0.0264 (0.0012) & -0.0320 (0.0013) & -0.0371 (0.0015) & -0.0406 (0.0015) \\
FD ($0.01$)                 & -0.0177 (0.0039) & -0.0260 (0.0037) & -0.0319 (0.0044) & -0.0385 (0.0046) \\
FD ($0.001$)                & -0.0216 (0.0124) & -0.0270 (0.0111) & -0.0275 (0.0139) & -0.0496 (0.0154) \\
\bottomrule
\end{tabular}
\end{table}
\vspace*{-1em}
\section{CONCLUSIONS}
\label{sec:conc}
In this paper, we propose a conditional Leibniz method by combining a recursive conditioning technique with a variant of the Leibniz divergence estimator from \shortciteN{ren2025stochastic}. The conditioning directly reduces variance. Moreover, we show that applying the divergence theorem to the Leibniz divergence estimator eliminates the LR term, thereby reducing the potential variance inflation w.r.t. the input dimension. We show how the conditional Leibniz estimator can be applied to sequential decision models whose decision rules are characterized by curves or regions in the state space. Simulation experiments based on an American call min-option model demonstrate the ability of the proposed method to provide low-variance and easy-to-implement estimators. 

There are several potential future research directions building on the results of this paper. An immediate extension is to incorporate the proposed estimator into gradient-based optimization algorithms. Also, for simplicity, we assume that the region $h(V,\theta)$ in the sample performance \Cref{eq:ori-perf} is a hyperrectangle. In practical applications, however, this region may have a more complicated shape. For example, in \Cref{fig:early-region}, the optimal early exercise region can be better approximated by adding a triangle to approximate the gray area, using two extra parameters; see, as another example, the control limit-type policy in \shortciteN{ren2025optimal}. New techniques are therefore needed to estimate surface integrals in such settings. More broadly, we hope to identify further applications of the proposed method; in particular, its discrete-time nature suggests potential applications to Bermudan-style options. Another important direction is to extend the recursive conditioning algorithm to more general stochastic processes. Although the methodology accommodates general Markov processes in principle, efficient sampling from the required conditional distributions may be challenging for non-Gaussian processes. We also notice that in \Cref{alg:example-leibniz}, each term $\mathcal{S}_i$ in the final estimator $\hat D$ corresponds to a different sample path and takes the form of a difference of payoff functions multiplied by a density term that acts as a weight. This suggests a natural direction for further variance reduction through importance sampling, e.g., by sampling different paths according to the magnitudes of their weights. Finally, notice that in \Cref{alg:example-leibniz}, although the sample paths used to estimate $\mathcal{S}_1$ and $\mathcal{S}_2$ are generated under different conditioning quantities, they are otherwise driven by i.i.d. Gaussian random vectors. This raises the question of whether the same data samples can be reused to generate different sample paths across algorithm replications to reduce data cost. It would be worthwhile to investigate this idea further in the context of data-driven optimization.

\section*{ACKNOWLEDGMENTS}
This work was supported in part by the U.S. National Science Foundation [Grants IIS-2123684, CMMI-2146067], and by the Natural Sciences and Engineering Research Council of Canada [Discovery Grant RGPIN-2018-05795].



\footnotesize

\bibliographystyle{wsc}

\bibliography{demobib}

\section*{AUTHOR BIOGRAPHIES}

\noindent {\bf XINGYU REN} received his Ph.D. from the Department of Electrical and Computer Engineering at the University of Maryland, College Park in 2026. He will be starting in September 2026 as a postdoctoral researcher at INSEAD working with Prof. Stephen Chick. His research interests include stochastic derivative estimation, stochastic optimization, and Markov decision processes. His e-mail address is \email{renxy@umd.edu}.\\

\noindent {\bf MICHAEL C. FU} holds the Smith Chair of Management Science in the Robert H. Smith School of Business, with a joint appointment in the Institute for Systems Research and an affiliate appointment in the Department of Electrical and Computer Engineering, at the University of Maryland, College Park. His research interests include stochastic gradient estimation, simulation optimization, and applied probability. He served as WSC2011 Program Chair and received the INFORMS Simulation Society's Distinguished Service Award in 2018. He is a Fellow of INFORMS and IEEE. His e-mail address is \email{mfu@umd.edu}.\\

\noindent
{\bf \MakeUppercase{Pierre L'Ecuyer}} is an Emeritus Professor in the
D\'epartement d'Informatique et de Recherche Op\'erationnelle,
Universit\'e de Montr\'eal, Canada.
He has published over 300 scientific articles and has developed software
libraries
and systems for random number generation and stochastic simulation.
He received the Lifetime Professional Achievement Award in 2020 and the Distinguished Service Award in 2011, both from the INFORMS Simulation Society. He is a Fellow of INFORMS. His e-mail address is
\url{lecuyer@iro.umontreal.ca}
and his website is
\href{https://www.iro.umontreal.ca/~lecuyer}
{https://www.iro.umontreal.ca/$\sim$lecuyer}.
\end{document}

%% file: wscbib.tex
\makeatletter
\let\@internalcite\cite
\def\cite{\def\@citeseppen{-1000}%
    \def\@cite##1##2{(##1\if@tempswa , ##2\fi)}%
    \def\citeauthoryear##1##2##3{##1 ##3}\@internalcite}
\def\citeNP{\def\@citeseppen{-1000}%
    \def\@cite##1##2{##1\if@tempswa , ##2\fi}%
    \def\citeauthoryear##1##2##3{##1 ##3}\@internalcite}
\def\citeN{\def\@citeseppen{-1000}%
    \def\@cite##1##2{##1\if@tempswa, ##2)\else{}\fi}%
    \def\citeauthoryear##1##2##3{##1 (##3)}\@citedata}
\def\citeA{\def\@citeseppen{-1000}%
    \def\@cite##1##2{(##1\if@tempswa , ##2\fi)}%
    \def\citeauthoryear##1##2##3{##1}\@internalcite}
\def\citeANP{\def\@citeseppen{-1000}%
    \def\@cite##1##2{##1\if@tempswa , ##2\fi}%
    \def\citeauthoryear##1##2##3{##1}\@internalcite}
\def\shortcite{\def\@citeseppen{-1000}%
    \def\@cite##1##2{(##1\if@tempswa , ##2\fi)}%
    \def\citeauthoryear##1##2##3{##2 ##3}\@internalcite}
\def\shortciteNP{\def\@citeseppen{-1000}%
    \def\@cite##1##2{##1\if@tempswa , ##2\fi}%
    \def\citeauthoryear##1##2##3{##2 ##3}\@internalcite}
\def\shortciteN{\def\@citeseppen{-1000}%
    \def\@cite##1##2{##1\if@tempswa, ##2\else{}\fi}%
    \def\citeauthoryear##1##2##3{##2 (##3)}\@citedata}
\def\shortciteA{\def\@citeseppen{-1000}%
    \def\@cite##1##2{(##1\if@tempswa , ##2\fi)}%
    \def\citeauthoryear##1##2##3{##2}\@internalcite}
\def\shortciteANP{\def\@citeseppen{-1000}%
    \def\@cite##1##2{##1\if@tempswa , ##2\fi}%
    \def\citeauthoryear##1##2##3{##2}\@internalcite}
\def\citeyear{\def\@citeseppen{-1000}%
    \def\@cite##1##2{(##1\if@tempswa , ##2\fi)}%
    \def\citeauthoryear##1##2##3{##3}\@citedata}
\def\citeyearNP{\def\@citeseppen{-1000}%
    \def\@cite##1##2{##1\if@tempswa , ##2\fi}%
    \def\citeauthoryear##1##2##3{##3}\@citedata}
%
%
%
\def\@citedata{%
    \@ifnextchar [{\@tempswatrue\@citedatax}%
                  {\@tempswafalse\@citedatax[]}%
}

\def\@citedatax[#1]#2{%
\if@filesw\immediate\write\@auxout{\string\citation{#2}}\fi%
  \def\@citea{}\@cite{\@for\@citeb:=#2\do%
    {\@citea\def\@citea{, }\@ifundefined
       {b@\@citeb}{{\bf ?}%
       \@warning{Citation `\@citeb' on page \thepage \space undefined}}%
{\csname b@\@citeb\endcsname}}}{#1}}%

%
\def\@citex[#1]#2{%
\if@filesw\immediate\write\@auxout{\string\citation{#2}}\fi%
  \def\@citea{}\@cite{\@for\@citeb:=#2\do%
    {\@citea\def\@citea{; }\@ifundefined
       {b@\@citeb}{{\bf ?}%
       \@warning{Citation `\@citeb' on page \thepage \space undefined}}%
{\csname b@\@citeb\endcsname}}}{#1}}%

%
\def\@biblabel#1{}
\makeatother



\newdimen\bibindent
\bibindent=0.0em
\def\thebibliography#1{\section*{\refname}\list
   {}{\settowidth\labelwidth{[#1]}
   \leftmargin\parindent
   \itemindent -\parindent
   \listparindent \itemindent
   \itemsep 0pt
   \parsep 0pt}
   \def\newblock{}
   \sloppy
   \sfcode`\.=1000\relax}